# Choosing a Knowledge Dissemination Technique

John Kingston
Health and Safety Laboratory
Buxton SK17 9JN
John.Kingston@hsl.gov.uk

## Abstract

Knowledge management has been described as 'getting the right knowledge to the right people in the right place at the right time'. Knowledge dissemination is a crucial part of knowledge management because it ensures knowledge is available to those who need it.

This paper reviews four well-known knowledge dissemination techniques. Each technique is classified according to a recently proposed classification scheme, and advice is given regarding when it is appropriate to use each technique.

## Introduction

Knowledge management (KM) has been defined as "*a conscious strategy for moving the right knowledge to the right people at the right time, to … improve organizational performance*" [1]. Knowledge dissemination – distributing knowledge to those who may need it – is therefore a crucial part of knowledge management.

A wide range of approaches to knowledge dissemination is available. It is important that organisations choose an approach that is appropriate for the knowledge being disseminated and for the organisation's structure, culture and business goals.

This paper begins by describing a recently published method that can be used to classify knowledge dissemination techniques. It then describes four knowledge dissemination techniques and when they should be used, including real life examples from the domain of health and safety. It concludes by summarising the pros and cons of the four categories within the chosen classification scheme.

This publication and the work it describes were funded by the Health and Safety Executive (HSE). Its contents, including any opinions and/or conclusions expressed, are those of the authors alone and do not necessarily reflect HSE policy.

# Classifying knowledge dissemination techniques

Milton [2] presents a four-way classification of systems for capturing and disseminating lessons learnt from past experience. It is presented here because it can also be applied to classify approaches for knowledge dissemination.

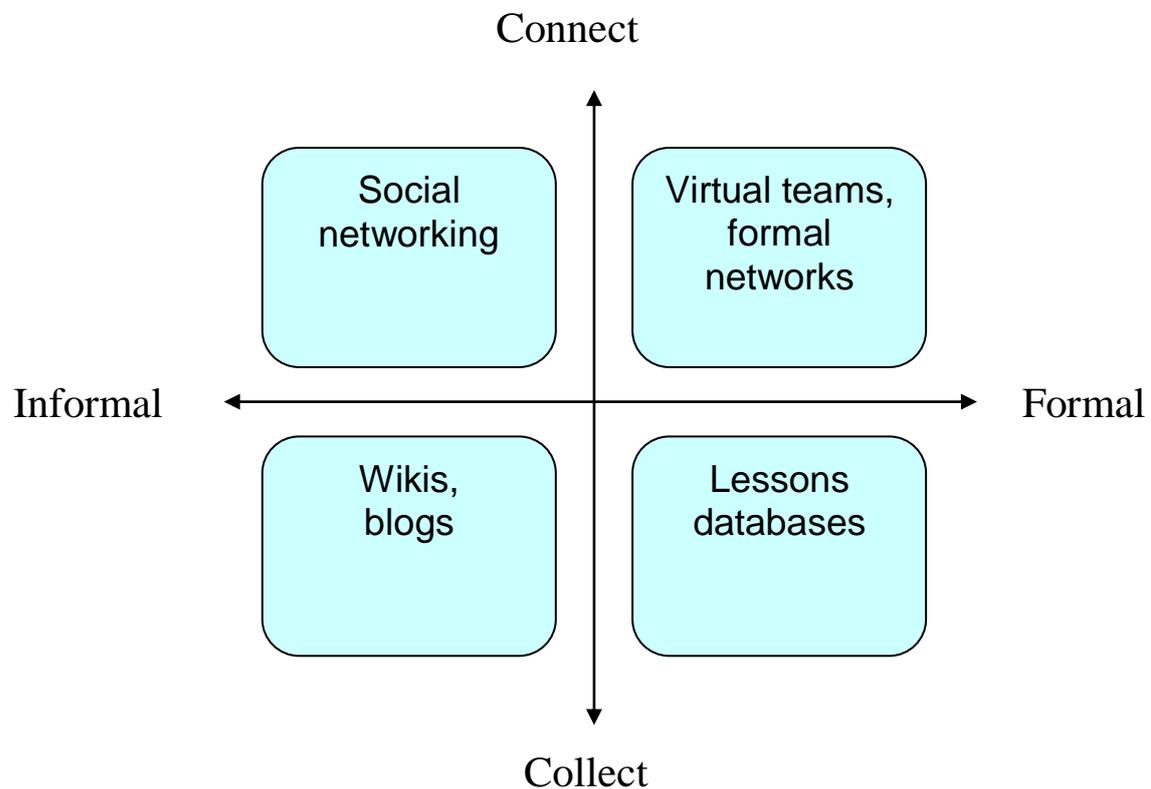

**Figure 1** Milton's classification of 'lessons learnt' approaches

Milton's classification uses two dimensions (see Figure 1). The first dimension distinguishes 'collect' approaches (in which knowledge is recorded or written into a repository) and 'connect' approaches (in which knowledge is communicated directly between individuals, whether verbally or through written messages). The second dimension distinguishes 'formal' and 'informal' approaches. By 'formal', Milton means "operating within a defined framework, or set of rules" [3], while 'informal' means "unmanaged and bottom-up"[3]. In practical terms, this often means that 'formal' techniques supply knowledge to users in a structured format, while informal techniques supply knowledge in conversational text.

This categorisation is powerful because it reflects one of the biggest philosophical debates in knowledge management research. This is the epistemological debate between 'cognitivists' and 'constructivists'. According to Heylighen [4], cognitivists believe that:

> … knowledge consists of models that attempt to represent the environment in such a way as to maximally simplify problem-solving. It is assumed that no model can ever hope to capture all relevant information, and even if such a complete model would exist, it would be too complicated to use in any practical way. […]. The model that is to be chosen depends on the problems that are to be solved. The basic criterion is that the model should produce correct (or approximate) predictions (which may be tested) or problem-solutions, and be as simple as possible."

Whereas the constructivist view is that:

1. Knowledge is not passively received either through the senses or by way of communication, but is actively built up by the cognising subject.
2. The function of cognition is adaptive and serves the subject's organisation of the experiential world, not the discovery of an objective ontological reality.

A phrase that is frequently used to summarise the constructivist view is that knowledge is a "justified personal belief" [5].

The practical implications of these views are that constructivists see knowledge as inseparable from an individual's mental model(s); this makes them favour 'connect' approaches to knowledge management, as it seems to them that anything 'collected' into a repository cannot be knowledge. The justification for this is that a repository is not animate, and is therefore not capable of holding a belief [6]; so anything in a repository must be information rather than knowledge. Constructivists also often prefer informal approaches, as there is no guarantee that the framework that accompanies any formal approach will match well with a learner's mental models. Cognitivists, however, aim to build models that represent the knowledge in someone's head but are separate from it (for a detailed methodology based on this approach, see [7]); they must therefore use 'collect' approaches. They also often prefer formal approaches, either because there are rules that describe how their models can be built, or because they are aiming to build a model whose structure reflects the expert's mental model(s). Therefore, it should be expected that there will be more proposed techniques (or at least, more enthusiasm for techniques) that fall into the 'informal connect' and 'formal collect' categories than in the other two categories..

Another KM debate that is informed by Milton's classification is the debate over whether tacit or explicit knowledge predominates in experts' heads, and what effect that should have on knowledge management techniques. Key arguments include the following:

- Dave Snowden, a well-respected KM practitioner argues that, because there are multiple factors in our environment and in our experience that influence us in ways we can never understand, we do not know what we know until we need to know it [8]. In other words, he argues that much of our knowledge is tacit until the context when it is needed arises. This is often taken to imply that 'collect' approaches to KM will be ineffective, as these approaches necessarily abstract knowledge away from its context.
- There are some situations where the context itself forms part of the necessary knowledge, but even the experts are unaware of this; perhaps the best known involved the original design of a specialised laser, in which the laser only functioned if a certain wire was below a certain length – no-one explicitly 'knew' this, but scientists who had viewed the original working set-up of the laser were able to re-create it and produce a working laser, whereas scientists who merely followed the circuit diagrams often could not do so [9]. In such situations, it is clear that the 'collected' knowledge is incomplete.

- The well-known 'knowledge creating cycle' proposed by Nonaka and Takeuchi [10] suggests that innovation occurs through a process that involves both tacit and explicit knowledge, but that the 'combination' step that creates new knowledge occurs when knowledge is tacit. This implies that 'collect' approaches will never fully support innovation, because knowledge that can be collected is always explicit; this implication is argued explicitly in [11].

In summary, any transfer of tacit knowledge is often deemed to require 'connect' approaches.

Cognitivists might reply to the above objections to 'collected' knowledge by arguing that context-dependent tacit knowledge can be made explicit, and then 'collected' by using realistic scenarios in knowledge capture sessions (e.g. see [12]). Dave Snowden himself accepts the need for some 'capture' of knowledge or information about the context in models [13]. Cognitivists would also argue that abstracting knowledge from its context is not only useful, it is essential if the important knowledge is to be clearly identified and its structure understood (cf. [9]). Regarding innovation, they would argue that for many business applications, it is not the creation of new knowledge that is important, but rather making existing knowledge available to those who need it in a usable form.

From a neutral standpoint, the most obvious point that emerges from these debates is that there appear to be two separate strands of practice that are both called 'knowledge management', but which are based on different theories, use different techniques, and support different business goals. Therefore, one of the keys to successful knowledge management in organisations is to understand the circumstances in which different techniques should be applied.

The remainder of this paper will examine four knowledge dissemination techniques; classify them into Milton's scheme; present examples of how they have been applied in a particular domain; and identify circumstances in which that technique is likely to be successful. It concludes with a review of the usefulness of Milton's classification scheme in selecting knowledge dissemination techniques.

## Examples of knowledge dissemination techniques

This section describes four knowledge dissemination approaches that were chosen because they were perceived to be commonly used in commerce, industry and government.

## Communities of practice – 'Informal Connect'

Communities of practice (COPs) have been defined as:

*"A group of individuals who share a common interest, a set of problems or a passion and who increase their knowledge and the understanding of these aspects through interpersonal relationships".* [14]

The key to using COPs as a KM dissemination technique is to encourage regular contact between COP members about knowledge need, available knowledge, and ideas. Contact can be either through face-to-face contact (in meetings and workshops, rather than one-to-one chats; the goal is for shared knowledge to be available to the whole community) or through internet-based discussion fora and/or other collaboration spaces.

Communities of practice are an 'informal connect' approach[1] to KM dissemination, according to Milton's categories; they connect those who need knowledge with those who have knowledge, and the knowledge that is expressed is usually in the form of conversations.

## Example: A Community of Practice in HSE

The UK Health and Safety Executive (HSE)'s mission is to prevent death, ill-health and injury to those at work and those affected by work activities. HSE's staff include (amongst others) inspectors, who inspect workplaces for health and safety risks or breaches of health and safety legislation; specialists, whose task is to maintain an in-depth up-to-date knowledge of a particular type of health and safety risk; and policy makers. HSE's staff are located in several offices, spread throughout Great Britain. HSE inspectors also spend considerable time visiting workplaces; if an incident has taken place at a workplace, they may be away from the office for several days at a time.

The Process Safety Community of Practice and Interest was originally conceived as a web based help desk in connection with setting out acceptable process safety standards for legal compliance. However, the ability to communicate widely and to interact with practitioners not only allowed the consultation and dissemination of new guidance and standards, but also allowed engagement of the wider community to consider particular technical problems and to agree an approach to take where clear decisions could not easily be made.

A clear advantage of the Process Safety community is that it engages a technical audience, with a broad base of skills and experience across wide and differing organisational boundaries. The community has been widened to include not only members of the process safety discipline, but also staff from mechanical engineering; electrical engineering; control and instrumentation; predictive specialists; and nuclear engineering.

One community member stated that this knowledge sharing forum provides a accessible conduit to "knowledge which is nascent in inspector's brains" and also provides a feeling that "you are part of a wider technical community, which can support you across organisational boundaries". It has been responsible for storing information; technical references; web-based links; material for developing new members of staff; and documenting approaches to particular technical issues. It has also been responsible for the development of new Regulators' Development Needs Assessment Competency Standards, which are implemented to support the performance review process for inspectors.

---

[1] It would be possible to turn COPs into an 'informal collect' approach by recording conversations and storing them in a central, searchable repository. This is rarely done, however, perhaps because wikis are a more effective method of collecting and organising knowledge from communities.

This community is successful because:

- It allows a widely dispersed group of differing technical specialists to communicate with each other across organisational boundaries;
- Community members can seek out knowledge and experience on a technical issue from a topic expert, who may otherwise remain unidentified within the organisation;
- The community can seek an independent opinion about a technical issue without formality and support an approach to take on challenging issues;
- Information is uploaded, downloaded, stored and communicated for the interest and benefit of all;
- The community's web site can be searched using technical search terms, which quickly generates a view about what is known within the community about a particular issue;
- The community promotes networking and problem solving;
- The community can generate new leads;
- The community reduces stress for inspectors who can otherwise feel very alone in making difficult technical decisions.

### When should communities of practice be used?

COPs work best when:

- Members have a shared history together, because establishing an effective COP requires a feeling of belonging to the community [15]. This factor highlights the importance of having face-to-face meetings of a community as well as an online discussion forum [16]. In the example above, the activities of the Process Safety community were supported by occasional face-to-face meetings;
- Geographically dispersed staff who have only occasional face-to-face contact need to work together, or to access each other's expertise [17];
- There is a need to obtain knowledge more quickly than it could be obtained from written sources [17];
- It is important to integrate new staff into a work community quickly [17] [18].
- There is an occasional need for new or innovative knowledge. New knowledge is often generated when a discussion of somebody's question leads not just to a solution, but also to an exchange of ideas about improved or new solutions  (as seen in the development of new standards by the process safety community; see also [17]).
- There is a possibility of unexpected or unpredictable interactions with users, such as demands for 'knowledge' that cannot be answered accurately without addressing the user's underlying assumptions[2]. In such cases, 'collected' knowledge must be accompanied by explanations of how the user's mental model needs to be modified to make use of the knowledge correctly. Humans who are having a conversation - or the written equivalent - are better than almost any collection of knowledge can be at diversifying responses, understanding context-based misapprehensions, and providing relevant explanations.

---

[2] An example is the infamous (and all too frequent) request to computer helpdesk technicians or computer repair shops for a replacement cup-holder, because the one supplied with the computer has broken. Before answering the query, it is highly advisable to explain to the user the difference between a cup-holder and a CD-ROM drive.

The size of a community should ideally be (as a rule of thumb) between 10 and 100 people. If the group is less than 10, it is likely that each of them will know all the others quite well, removing some of the need for a COP, and that any online forum will be accessed only rarely, thus reducing its usefulness. As for the upper limit, most literature on KM implicitly assumes that large knowledge-sharing groups are more effective and efficient than small groups [19] [20] because *"there is a big enough group to create a useful database of information"* [21]. However, Butler [22] argued that large numbers could negatively affect a community, as having too many members can make it difficult to converge on a joint enterprise; interactions between any subset of members may be occasional and unsustainable; and shared repertoires may be difficult to build, maintain, and understand. There is also the consideration that a large community that asks many questions might create unacceptable demands on the time of the community's experts. Voelpel *et al* [23] also discovered that smaller groups are more likely to respond or engage in the act of knowledge sharing, while still providing a comparably high level of knowledge. Voelpel *et al* propose that COP group sizes should be limited to around 100 members.

## Knowledge Portals – 'Formal Connect'

Knowledge Portals (KP) are IT systems, often with a web page as a user interface, that provide a single point of access to key information and knowledge resources [24]. The information supplied can be customised according to user requirements. They can bring together information from internal and external sources that is relevant to an individual or a group. Commonly-used sources include corporate message boards, document search results, internal databases, specialised websites, or news channels [25].

In technical terms, KP are 'middleware': they provide a mediating layer between the information applications and the individuals using them (see Figure 2).

KP were first developed as organisations sought to unify information access and improve the management of information resources by trying to replicate the popularity and success of public domain, web-based portals such as Yahoo and Google [26] [27]. The KP approach can be seen as a natural evolution of intranets and groupware solutions into a common information infrastructure [28].

In Milton's classification, knowledge portals are 'formal connect' systems. They connect users with other sources of knowledge, rather than storing knowledge in a repository. The formal design of the knowledge portal acts as a structure for the knowledge contained within it.

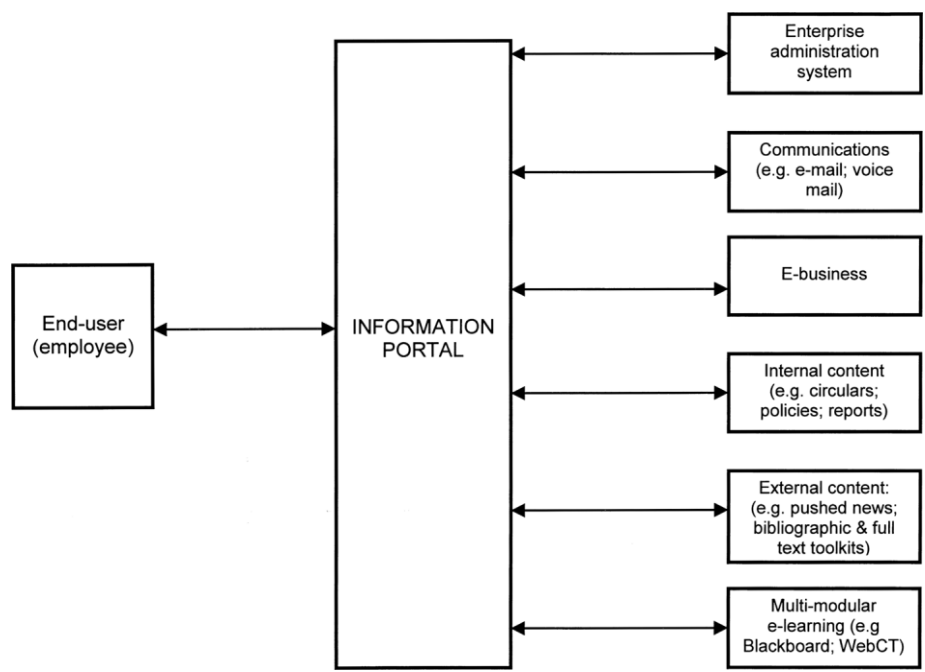

**Figure 2** The portal as 'middleware' [29]

## *Example: A Knowledge Portal in HSE's Chemicals Regulation Directorate*

HSE's Chemical Regulation Directorate (CRD) has the task of approving chemical substances (including pesticides) for use, and recording that approval with a Europe-wide database. CRD needs to access relevant regulations, associated guidance and studies that have been carried out on the substance. There is a great deal of guidance available for pesticides; the regulations, the guidance and other necessary information are stored on a variety of databases and websites. Some information is buried several levels down in European Commission websites, which are not always easy to navigate. Furthermore, requests for approvals are submitted electronically in a structured format, and records of requests (and responses) must be stored on a database run by the European Commission in a similar structured format.

CRD introduced a knowledge portal to address the challenge of providing its staff with access to relevant technical information on pesticides. When the user selects a substance from the master list of substances (there are about 200 substances in total), a page of further links is displayed, including the background to decision-making and issues discussed. Staff from each section who have received brief training from the IT department are able to create customised portals that provide extra information relevant to that section.

This knowledge portal approach is successful because:
- it can provide users with links to all and only the information sources relevant to their work;
- it gives users direct access to hard-to-find web pages;

- it is not technically difficult to set up, so users can be trained to set up their own systems;
- by eliminating intermediate pages and by caching frequently accessed pages, it reduces 'click-through' time and produces considerable improvement in internet efficiency.

Another indirect advantage is that the file-sharing facilities of the underlying software (Microsoft SharePoint) mitigate the potential risks to reputation and information security associated with sharing knowledge between staff at CRD's multiple sites.

## *When should knowledge portals be used?*

Knowledge portals should be used:
- Where there are several key IT systems through which staff repeatedly need to access specific information [30];
- Where frequently used IT systems are difficult to navigate, so staff would benefit from a customised interface to these systems [31];
- When there is a need for knowledge to be drawn together from, or distributed to, loosely coupled or disconnected knowledge sources e.g. two different databases [32];
- Where staff would benefit not only from access to key IT systems but also improved access to collaborative tools in order to share information and knowledge with each other [32] [33] [34] [35].

# Knowledge Codification - 'Formal collect'

Codified knowledge is defined here as knowledge that has been re-formatted and/or had additional indexing features added to it. The purpose of this is to make the knowledge more understandable and accessible, whether by a person or by a computer program.

There are several methods of knowledge codification requiring different degrees of modification of the knowledge in question.

### Re-formatting: small modifications
Codification methods that require only small modifications include drawing a diagram to represent a process that was originally described in words (creating a 'knowledge map'); re-arranging the order of sections in a document; or abbreviating long paragraphs of text as checklists – for example, distilling the records of 'lessons learnt' exercises into a list of key lessons. The codified knowledge that results from these processes is usually suitable for user manuals or databases; an example can be found in [36].

### Re-formatting: large modifications
Larger modifications to knowledge are usually required if the knowledge is to be encoded into an IT system. Knowledge that has been captured as stories or in structured interviews may be searched in order to extract knowledge that fits certain formats (e.g. rules of the form *If X is true then Y is true*[3]). Alternatively, one or more knowledge maps based on the text may be produced, with the final map(s) being used as a specification for an IT system [37]. Databases may similarly be analysed for associations between data entries that can be transformed into rules or a relationship map [38].

**Indexing: small modifications**
Indexing schemes can also be small-scale or large-scale. Small-scale indexing schemes include creating an index to a book or document, or creating hyperlinks within a document to other parts of the document.

**Indexing: large modifications**
Large-scale indexing includes:
- identifying and linking a list of key features to 'knowledge items' in the document [39];
- extracting the 'knowledge items' and positioning them within a newly created or existing taxonomy or other classification scheme [40]; and
- running analyses on text in documents in order to group them with other documents with similar keywords [41].

Re-formatted knowledge is most often used to make guidance or procedures available to its intended users. If it is in the form of a manual or guidance document, it can be distributed as a readable document, whether on paper or electronically. If the knowledge is encoded into an IT system, the system will typically ask the user some questions about the features of a current situation or problem, and will use its encoded knowledge to suggest guidance for the situation or a solution to the problem.

Indexed knowledge is used to support searching of a large collection of knowledge. It is most often used to support 'lessons learnt' processes, i.e. to search a large collection of past cases/incidents/projects for those that are relevant to the current situation. The enquirer can extract and apply the appropriate lessons.

Supporting software is available for all the codification activities listed above:

- Small-scale re-formatting of knowledge can be supported by computer programs such as computer-aided software engineering tools or mind mapping software;
- Expert systems are computer programs that encode expert knowledge that has been through large-scale reformatting;
- Many technologies that support small-scale indexing of knowledge are incorporated within popular desk-top publishing packages;
- Once large-scale indexing has been performed, case-based reasoning is a computer technology that scans a database of past problems and solutions, and uses indexed key features to match a current case against the most relevant past case(s).

---

[3] For example, when capturing knowledge about eligibility for welfare benefits under English law, one rule might be: "IF you have a spouse or partner who is working full-time THEN you are not eligible for benefits X and Y".

 

Knowledge codification requires that all knowledge is collected into a repository before being represented in a codified form; and the codification process always requires organising the knowledge into some kind of structure, framework or model. In Milton's classification, knowledge codification techniques can therefore be considered to be the epitome of 'formal collect' systems.

## *Example: Expert system in Occupational Hygiene*

EASE (Estimation and Assessment of Substance Exposure [42; 43; 44]) is an expert system built for HSE for assessing workplace exposure to potentially hazardous substances.

The objective of the system is to enable assessors (inspectors or specialists with expertise in occupational hygiene) to produce an estimate of the extent to which workers will be exposed to a new or existing hazardous substance. It achieves this by guiding the user through a decision tree, with conclusions that are drawn from an analysis of HSE's National Exposure Database. The system's completeness was validated by running exhaustive tests that followed every possible path through its decision tree, and then checking the accuracy of its outputs.

The system helps assessors to remember all the necessary questions to ask when considering a new industrial process, and then estimates an exposure level from their answers. The system also incorporates backtracking facilities to allow errors (e.g. if the user supplies incompatible answers to two questions) to be corrected and 'What If' analyses to be performed.

EASE has now been superseded by other expert systems, notably the Advanced Reach Tool [45], but its lasting benefit to HSE has been its influence on approaches to exposure assessment within the European regulatory framework.

## *When should codified knowledge be used?*

Codification of knowledge is useful:
* When there are many potential users of the disseminated knowledge or where the potential cost savings are very large. Codified knowledge is expensive to produce compared with other knowledge dissemination approaches, especially if large-scale modifications are made, so there needs to be a large user base to make it cost-effective. See [46] for an example of an expert system that saved a large procurement department, with a few hundred staff, an estimated £30 million per year;
* Where there is a strong desire for a group of users to use a standard form of knowledge and/or a standard approach to solve a particular problem;
* Where a task can be performed better by an expert system than by a human expert because of time constraints. For example, the task of scheduling (or dynamic rescheduling) of aeroplanes by a major airline, in order to ensure that aeroplanes arrive at maintenance hubs just before their regular maintenance check-ups are due, can be performed better by an expert system than by a human expert [47];

- Where the knowledge concerns a specific task that includes several decision points;
- Where knowledge consists of an overview of the organisation, including topics such as who has the authority to make decisions, or which departments carry out which procedures. Knowledge mapping can provide a guide to this type of knowledge, and also to when, why, or with what resources tasks are performed (see [48]);
- Where the specific task being considered is one to which all possible solutions can be listed. This is a 'rule of thumb', because it is possible to use codified knowledge to support tasks where it is not practical to list all solutions (such as the airline scheduling task described above), but it is usually less cost-effective to support such tasks;
- When it is important to be sure that the knowledge being supplied by a dissemination technique is complete (verification) and accurate (validation). This is only possible when using 'collect' approaches, and is easier with 'formal collect' approaches, because the restructuring of the collected knowledge often highlights issues relating to verification or validation. An example of such an issue is the possibility, identified in the EASE system, that users might give answers to two questions that are incompatible with each other.

# Apprenticeship/mentoring and training – 'Informal connect/Formal connect'

The apprenticeship model is one of the oldest forms of learning, in which a novice obtains practical experience by working with and being instructed by a skilled craftsman, artisan, or tradesman. A genuine apprenticeship involves a long-term commitment that is designed not merely to enable an apprentice to achieve a threshold level of acceptability, but to prepare the newcomer to function as a professional member of that trade [49]. Mentoring works on the same principles as apprenticeship, but may be less formal, or of shorter duration: for example, job rotation or job swaps often require setting up a mentoring relationship.

Features of successful apprenticeships include:

- Recognition of achievements;

- Entry into a community of practice[4], including social initiation into the rules of the community;

- Purposeful learning by carrying out an authentic task;

- Learning through practice, in the context within which the task is most functional and useful;

- Tasks that are graded and progressive;

- Initiation of apprentices being seen as important by the skilled staff, even when this process might hinder the speed of work.

---

[4] This refers to a 'community of practice' in the sense of a group of people with similar skills and interests.

Training courses are often used to accompany apprenticeship/mentoring, so that staff learn formally from the trainer and then put their training into practice in the role of apprentices.

Training is a 'formal connect' approach, in Milton's terms – the trainees are given the chance to connect with a teacher who conveys structured knowledge to them. Apprenticeship/mentoring is an 'informal connect' approach, though with a few formal aspects e.g. in the progression of tasks that is organised, and in the recognition of achievements.

## Example: Mentoring in HSE's inspector training

HSE's training programme for newly recruited inspectors can be considered to operate on a training/mentoring model:
* The training programme is based upon sequenced stages and includes a mixture of on the job experience and formal training, designed to develop the skills and knowledge required of an independent inspector.
* The mentoring aspects are provided to new inspectors through joint visits to workplaces with experienced inspectors from their own and other groups, as well as observing specific procedures such as prosecutions, tribunals and inquests. The new inspector's role progresses from involved observer to lead inspector, and then to independent inspector.

## When should apprenticeship/mentoring be used?

Apprenticeship/mentoring can be seen as the default approach to knowledge dissemination; it will work in nearly all situations, but it is more time-consuming than other methods. It takes a long time, typically 2-5 years, before the apprentice is fully experienced. It also requires quite a lot of the expert's time for each learner; it is not practical for an expert to pass on his knowledge to more than a handful of apprentices while still performing his regular tasks. The use of formal training reduces the latter problem, but apprenticeship/mentoring cannot compete for efficiency with other knowledge dissemination techniques, which can reach many learners simultaneously.

Apprenticeship/mentoring is particularly useful for disseminating tacit knowledge, or for tasks that require knowledge of graphics or diagrams, perception, or physiology (i.e. coordinated muscle movements). An example of perceptual, geometric and physiological knowledge can be found in the rail industry: the (now largely obsolete) task of 'wheel tapping' involved striking the wheel of a railway engine or carriage with a long-handled hammer and listening to the resulting sound to determine if the wheel was cracked or damaged. The primary skill used by wheel tappers was perceptual (identifying 'incorrect' sounds); there were also secondary skills in knowing where and how hard to tap the wheel (geometric and physiological knowledge).

# Conclusion

This conclusion highlights the pros and cons of knowledge dissemination systems within each of Milton's four categories. These features should be taken into consideration when choosing which type of system is appropriate.

**Informal connect** techniques are typically the easiest to set up, and (if the experts are readily available) quick to use. They allow knowledge seekers to hold conversations with experts, which is important if the knowledge must be explained in more than one way before the seeker can integrate the knowledge into his mental model. These conversations can also be a source of unexpected knowledge for seekers, or even of creativity or innovation, especially if more than one expert is involved.

It was predicted earlier in this paper that there would be more, or more enthusiasm for, 'informal connect' than other types of technique. This seems to be the case: there are several techniques that are designed to allow people to share knowledge verbally and informally, albeit within a structured context. These techniques include peer assists [50]; storytelling [51; 52]; 'an audience with' (a highly interactive seminar); and social media. The social media model in which someone makes a statement and others choose to comment on it is similar to the approach of an online community of practice, if the topic of the statement is a request for knowledge or an offer of knowledge.

There are several arguments in the literature that suggest that informal connect techniques are favoured by knowledge seekers. Indeed, Cross and Sproull [53] found that, given a choice of approaches, users will often choose to obtain knowledge by 'connecting' with nearby experts, even when collected knowledge is easily available and of high quality. The reason is that conversations with more experienced individuals often supply staff with more information than just the answer to the question. They may also get meta-knowledge (where to go to get more information on the issue, or conversely where not to go because a certain report is outdated); problem reformulation (when the expert suggests a different way to look at the problem or issue, or highlights potentially unforeseen consequences of actions); validation (assurance that the approach the seeker is taking is on course, and consequent appreciation); and legitimising (an expression of approval by a person in authority or with known expertise, which the seeker can then use to influence others). Milton [4] also suggests that informal connect systems are popular because their informality makes them easy to use, and because users have the freedom to introduce new discussion topics in their own words rather than trying to understand the system's language or classification.

However, the potential for new discussions to open up is also one of the disadvantages of informal connect systems; in Milton's words, they can end up more as gossip than an exchange of knowledge. Another disadvantage is that the knowledge is usually not stored for the benefit of later users who might need it; there is a risk that only those who use the system at the time the request is answered will benefit. It is also difficult for the organisation to verify or standardise the knowledge that is shared.

 

Informal connect systems are also relatively demanding of the time of experts, which becomes a problem if the number of people seeking their knowledge exceeds 100 or so.

**Formal connect** systems typically require basic IT skills from developers and users. They are mainly useful when there are a number of information sources (human or IT) that need to be consulted and the resulting information amalgamated to solve a problem; or where there are information sources that need to be consulted regularly because their content is regularly updated. These information sources may be in-house sources that consist of collected knowledge; for example, Milton argues that formal connect systems are ideal for sharing 'lessons learnt' in areas of complex or context-specific need, or for accessing knowledge about topics that are rapidly changing. Formal connect techniques may therefore be most useful when used in conjunction with collect approaches.

Formal connect systems usually reduce effort for the knowledge users, but do not necessarily reduce effort for knowledge providers.

**Informal collect** approaches arguably include all forms of publishing[5], but a more practical view restricts 'informal collect' knowledge dissemination techniques to material that is 'published' on a voluntary or as-needed basis in order to address a perceived lack of available knowledge. Techniques in this category include blogs and wikis. Published text is usually easy to understand, and often provides background context as well. However, text can sometimes be hard to search or to query, particularly in cases where multiple terms can be used to describe similar concepts. Semantic tagging (e.g. within semantic blogs [56]) can reduce these difficulties.

Collect approaches are always more laborious to implement and to maintain than connect approaches, as the knowledge has to be written down ('captured'), and the user must then search it for knowledge relevant to their needs. However, the 'capture'process can be shared between many people in systems such as wikis. Perhaps the biggest advantages of 'collect' approaches within organisations are that the knowledge now resides in a record, and that record belongs to the organisation rather than to an individual. The fact that the knowledge is recorded means that the knowledge can be verified and validated and the results can be supplied to users as 'standard' knowledge. However, a corresponding disadvantage of voluntary approaches to knowledge sharing is that the body of collected knowledge is likely to be very incomplete; Milton argues that voluntary wikis draw on only about 2-3% of all available knowledge about a topic.

**Formal collect** systems require the most effort to establish, as the knowledge has to go through three distinct stages: capture; structuring; and making the

---

[5] Many organisations have libraries of internal reports, or access to collections of publications on the Web, that contain valuable knowledge. The primary technique at present for searching such repositories for knowledge is text mining (as introduced in [41]; see [54] and [55] for some examples). However, since text mining is a technique for searching for knowledge rather than for disseminating knowledge, it is beyond the scope of this paper.

  

knowledge comprehensible for dissemination. However, this very process subjects the knowledge to greater analysis than in any of the other techniques, which brings advantages if the knowledge needs to be verified, validated, standardised, or made easier to understand. The way in which users access or make use of this knowledge can also be controlled quite tightly (if desired) by building it into IT systems. The ultimate in control is the expert system, which is unique amongst knowledge dissemination techniques in that its primary purpose is not to supply knowledge to users to enable them to make decisions; instead, it directs queries to the users, makes its own decisions, and then produces an answer. For users who want to learn how decisions were made, or to decide whether the system is trustworthy, appropriate justifications are usually made available. These systems are therefore most suitable for tasks where there are a large number of users with low levels of expertise.

Alternatively, IT-based 'formal collect' systems can utilise the power of computing to automate tasks that are too complicated for an expert to perform optimally; these are usually tasks such as configuration, design, planning or scheduling that follow a set of knowledge-based principles with a set of constraints, but have a very large number of possible solutions.

The biggest difficulty that formal collect systems have is being applicable to a wide range of problems. The need to follow a structured dialogue may occasionally make it difficult for users to enter the content into the system that they want to enter; and if any knowledge is required that goes beyond the knowledge stored in the system, then a formal collect system is simply unable to supply it or support a search for it.

It was predicted earlier that formal collect approaches would outnumber, or at least engender more enthusiasm, than informal collect or formal connect approaches. In practice, this prediction seems to be mitigated by users' natural preference for informal connect approaches (as identified by Cross and Sproull) and by the effort required to design and distribute formally collected knowledge. However, in circumstances where formal collect approaches are justfiied, they can provide organisations with very large savings through improved efficiency and through more effective decision making.

Disclaimer: This publication and the work it describes were funded by the Health and Safety Executive (HSE) and the Health and Safety Laboratory (HSL). Its contents, including any opinions and/or conclusions expressed, are those of the author alone and do not necessarily reflect HSE or HSL policy.

# References

[1] O'Dell C, Grayson CJ. *If we only knew what we know: the transfer of internal knowledge*. The Free Press; 1998

[2] Milton N. *The Lessons Learnt Handbook: Practical Approaches to Learning from Experience*. Chandos Publishing; 2010

[3] Milton N. *The Polarity of KM Ideologies*. http://www.nickmilton.com/2009/10/polarity-of-km-ideologies.html. Accessed February 2011

[4] Heylighen F. *Epistemological Constructivism*. http://pespmc1.vub.ac.be/EPISTEMI.html. Downloaded March 2011

[5] Alavi M. and Leidner D. *Knowledge Management Systems: Issues, Challenges, and Benefits*. Communication of AIS, Vol. 1, Article 14: 1999

[6] Goodwin, S.C., *Formal Knowledge Sharing in Medium-to-Large Organizations: Constraints, Enablers and Alignment*. Thesis, (PhD), University of Bath. 2009

[7] Schreiber G., Akkermans H., Anjewierden A., de Hoog R., Shadbolt N., van de Velde W., Wielinga B. *Knowledge Engineering and Management: The CommonKADS Methodology*, The MIT Press, Cambridge, MA, 2000

[8] Snowden,DJ and Boone ME. *A Leader's Framework for Decision Making*. Harvard Business Review, November 2007

[9] Collins HM and Harrison R. *Building a TEA Laser: The Caprices of Communication*, Social Studies of Science, 5, 441-50: 1975

[10] Nonaka I and Takeuchi H. *The knowledge creating company: how Japanese companies create the dynamics of innovation*, New York: Oxford University Press. 1995

[11] Swan, J. Newell, S. Scarbrough, H. and Hislop, D. *Knowledge management and innovation: networks and networking*. Journal of Knowledge Management, 3,4, 262-275: 1999

[12] van Someren MW, Barnard YF, Sandberg JAC. *The Think Aloud method: A practical guide to modelling cognitive processes*. Department of Social Science Informatics, University of Amsterdam. Academic Press, London, 1994

[11] Snowden D., *Sensemaking*, www.cognitive-edge.com, downloaded June 2011

[14] Wenger E, McDermott R, Snyder WM. *Cultivating Communities of Practice: A Guide to Managing Knowledge*. Harvard Business School Press, Cambridge, USA; 2002

[15] Akkerman S, Petter C, De Latt M. *Organising communities of practice: facilitating emergence*. Journal of Workplace Learning, 20 (6): 383-399, 2008

[16] Dube L, Bourhis A, Jacob R. *The impact of structuring characteristics on the launching of virtual communities of practice*. Journal of Organisational Change Management. 18 (2): 145-166, 2005

[17] Ardichvili A, Page V, Wentling T. *Motivation and Barriers to Participation in Virtual Knowledge-Sharing Communities of Practice*, OKLO 2002 Conference, Athens, Greece. 2002

[18] Rollag K, Parise S, Cross R. *Getting New Hires Up to Speed Quickly*, MIT/Sloan Management Review, Winter Issue: 35-44, 2005

[19] Mahnke V. *The Economies of Knowledge-Sharing: Production- and Organization Cost Considerations*. 6th conference of the Danish Research Unit for Industrial Dynamics, Winter 1999

[20] Hansen JR, Nohria N, Tierney T. *What's Your Strategy for Managing Knowledge?* Harvard Business Review, March-April 1999

[21] Cabrera A, Cabrera EF. *Knowledge-sharing dilemmas*. Organization Studies 23(5):687-710. 2002

[22] Butler T, Murphy C. *Implementing Knowledge Management Systems in Public Sector Organisations: A Case Study of Critical Success Factors,* 15th European Conference on Information Systems, St. Gallen, Switzerland, 12: 612-623. 2007

[23] Voelpel SC, Eckhoff RA, Forster J. *David against Goliath? Group size and bystander effects in virtual knowledge sharing*. Human Relations, 61 (2): 271-295. 2008

[24] Reneker MH, Buntzen JL. *Enterprise knowledge portals: two projects in the United States*. Department of the Navy. Electronic Library 18(6): 392. 2000

[25] Cloete M, Snyman R. *The enterprise portal - is it knowledge management?* Aslib Proceedings 55(4): 234-242. 2003

[26] Mack R, Ravin Y, Byrd RJ. *Knowledge portals and the emerging digital knowledge workplace*. IBM

Systems Journal 40(4), 925. 2001

[27] Harris K, Phifer G, Hayward S. *The enterprise portal: Is it knowledge management?* Gartner; SPA-08-8978, 2 pages, available at www.gartner.com, 1999. Accessed January 2011

[28] Watson J, Fenner J. *Understanding Portals*. Information Management Journal 34(3), 18-22. 2000

[29] Van Brakel P. *Information portals: a strategy for importing external content.* The Electronic Library 21(6): 591-600. 2003

[30] Daniel E, Ward J. *Enterprise portals: Addressing the organisational and individual perspectives of information services*. 13th European Conference on Information Systems. Regensburg, Germany, May 26-28 2005

[31] Mack R, Ravin Y, Byrd RJ. Knowledge portals and the emerging digital knowledge workplace. IBM Systems Journal 40 (4): 925-955, 2001

[32] van Baalen P, Bloemhof-Ruwaard J, van Heck E. *Knowledge sharing in an emerging network of practice: The role of a knowledge portal.* European Management Journal 23 (3): 300-314, 2005

[33] Dias C. *Corporate portals: a literature review of a new concept in Information Management.* International Journal of Information Management 21 (4):269-87, 2001

[34] Firestone JM. *Enterprise Information Portals and Knowledge Management*. Boston, MA: KMCI Press/Butterworth Heinemann; 2003

[35] Detlor B. *The corporate portal as information infrastructure: towards a framework for portal design*. International Journal of Information Management 20 (2): 91-101, 2000

[36] Johnson AM, *Using Concept Mapping to Facilitate Students Understanding Software Testing*, http://www.testingeducation.org/conference/wtst4/AMJohnsonUsing%20Concept%20Mapping%20to%20Teach%20Software%20Testing%20(2).doc, accessed January 2011

[37] Kingston JKC, *Designing Knowledge Based Systems: The CommonKADS Design Model*, Knowledge Based Systems Journal, 11 (5-6): 311-319. 1998

[38] Langley P, Simon HA, *Applications of Machine Learning and Rule Induction,* Communications of the ACM, 38(11), November 1995

[39] Watson I, Marir F. *Case based Reasoning: A Review*. Knowledge Engineering Review, 9(4): 327-354, 1994. Also available at http://www.ai-cbr.org/classroom/cbr-review.html. Accessed January 2011

[40] Chaudhry AS. *Assessment of taxonomy building tools*, The Electronic Library, 28(6): 769-788. 2010

[41] Marinai S. *Introduction to Document Analysis and Recognition*, Studies in Computational Intelligence 90: 1-20. 2008

[42] Tickner J, Friar J, Creely KS, Cherrie JW, Pryde DE, Kingston JKC. *The Development of the EASE Model*. Annals of Occupational Hygiene, 49(2): 103-110. 2005

[43] Kingston JKC. *Building a KBS for Health and Safety Assessment*. In: Applications and Innovations in Expert Systems IV, proceedings of BCS Expert Systems'96, Cambridge, 16-18 December 1996. SGES Publications; 1996

[44] Creely KS, Tickner J, Soutar AJ, Hughson GW, Pryde DE, Warren ND, Rae R, Money C, Phillips A, Cherrie JW. (2005) *Evaluation and further development of EASE model 2.0.* Annals of Occupational Hygiene, 49(2): 135‑46; 2005.

[45] Tielemans E, Warren NE, Fransman W, Van Tongeren M, Mcnally K, Tischer M, Ritchie P, Kromhout H, Schinkel J, Schneider T, Cherrie JW. *Advanced REACH Tool (ART): Overview of Version 1.0 and Research Needs*. Annals of Occupational Hygiene, 55(9): 949-956; 2011. See also http://www.advancedreachtool.com, accessed March 2012

[46] Power R, Reynolds S, Kingston J, Harrison I, Macintosh A, Tonberg J. *Expert Provisioner: A Range Management Aid*, Knowledge Based Systems Journal, 11 (5-6): 339-344. 1998

[47] Smits S, Pracht D. *MOCA—A Knowledge-Based System for Airline Maintenance Scheduling*. In Proceedings of the Innovative Applications of Artificial Intelligence Conference. AAAI; 1991

[48] Kingston JKC. *Conducting Feasibility Studies for Knowledge Based Systems*. Knowledge-Based Systems, 17 (2-4): 157-164, May 2004

[49] Douglas RL. *Arbitrators, apprentices, and arbitration.* The Arbitration Journal 37(3): 46-51. 1982

[50] Knoco Ltd. *Knowledge Management Processes: Peer Assist*. http://www.knoco.com/peer-assist-page.htm. Accessed February 2012.

[51]   Denning S. *Telling Tales*, Harvard Business Review, May 2004

[52]   Perrett R, Borges MRS, Santoro FM. *Applying Group Storytelling in Knowledge Management*. In: Groupware: Design, Implementation and Use. Lecture Notes in Computer Science, 2004

[53]   Cross R, Sproull L. *More Than an Answer: Information Relationships for Actionable Knowledge*. Organization Science. 15 (4): 446-462, 2004

[54]   Kiryakov A, Bishop B, Ognyanoff D, Peikov I, Tashev Z, Velkov R. *The Features of BigOWLIM that Enabled the BBC's World Cup Website.* Proceedings of the Workshop on Semantic Data Management at VLDB 2010, Singapore. http://ceur-ws.org/Vol-637/paper6.pdf. Sep 17 2010

[55]   Doms A, Schroeder M. *GoPubMed: exploring PubMed with the Gene Ontology*. Nucleic Acids Research, 33 suppl. 2: W783-W786. Oxford University Press. http://nar.oxfordjournals.org/content/33/suppl_2/W783.long, 2005. Accessed February 2012.

[56]   Cayzer S. *Semantic Blogging and Decentralized Knowledge Management*, Communications of the ACM, 47(12). December 2004